\documentclass[twocolumn]{aastex631}

\begin{document}

\title{Turbulence in Simulated Local Cluster Analogs: one-to-one comparisons between SLOW and XRISM/Hitomi}

\correspondingauthor{Frederick Groth}
\email{fgroth@usm.lmu.de}

\author[0000-0002-1687-6774]{Frederick Groth}
\affiliation{Universit{\"a}ts-Sternwarte, Fakult{\"a}t für Physik, Ludwig-Maximilians-Universit{\"a}t M{\"u}nchen, Scheinerstr. 1, 81679 M{\"u}nchen, Germany}

\author[0000-0002-0796-8132]{Milena Valentini}
\affiliation{Dipartimento di Fisica, Università degli Studi di Trieste Via Alfonso Valerio 2, 34127 Trieste, Italy}
\affiliation{INAF - Astronomical Observatory of Trieste, via Tiepolo 11, I-34131 Trieste, Italy}
\affiliation{INFN, Instituto Nazionale di Fisica Nucleare, Via Valerio 2, I-34127, Trieste, Italy}
\affiliation{ICSC - Italian Research Center on High Performance Computing, Big Data and Quantum Computing, via Magnanelli 2, 40033, Casalecchio di Reno, Italy}

\author{Benjamin A. Seidel}
\affiliation{Universit{\"a}ts-Sternwarte, Fakult{\"a}t für Physik, Ludwig-Maximilians-Universit{\"a}t M{\"u}nchen, Scheinerstr. 1, 81679 M{\"u}nchen, Germany}

\author{Stephan Vladutescu-Zopp}
\affiliation{Universit{\"a}ts-Sternwarte, Fakult{\"a}t für Physik, Ludwig-Maximilians-Universit{\"a}t M{\"u}nchen, Scheinerstr. 1, 81679 M{\"u}nchen, Germany}
\affiliation{European Southern Observatory, Karl Schwarzschildstrasse 2, 85748, Garching bei M\"unchen, Germany}

\author{Veronica Biffi}
\affiliation{INAF - Astronomical Observatory of Trieste, via Tiepolo 11, I-34131 Trieste, Italy}
\affiliation{IFPU – Institute for Fundamental Physics of the Universe, Via Beirut 2, I-34014, Trieste, Italy}

\author{Klaus Dolag}
\affiliation{Universit{\"a}ts-Sternwarte, Fakult{\"a}t für Physik, Ludwig-Maximilians-Universit{\"a}t M{\"u}nchen, Scheinerstr. 1, 81679 M{\"u}nchen, Germany}
\affiliation{Max-Planck-Institut f{\"u}r Astrophysik, Karl-Schwarzschild-Stra{\ss}e 1, 85741 Garching, Germany}

\author[0000-0002-2307-2432]{Jenny G. Sorce}
\affiliation{Univ. Lille, CNRS, Centrale Lille, UMR 9189 CRIStAL, 59000 Lille, France}
\affiliation{Université Paris-Saclay, CNRS, Institut d'Astrophysique Spatiale, 91405 Orsay, France}

\begin{abstract}

The XRISM Resolve X-ray spectrometer allows to gain detailed insight into gas motions of the intra cluster medium (ICM) of galaxy clusters. Current simulation studies focus mainly on statistical comparisons, making the comparison to the currently still small number of clusters difficult due to unknown selection effects. This study aims to bridge this gap, using simulated counterparts of Coma, Virgo, and Perseus from the SLOW constrained simulations. These clusters show excellent agreement in their properties and dynamical state with observations, thus providing an ideal testbed to understand the processes shaping the properties of the ICM. 
We find that the simulations match the order of the amount of turbulence for the three considered clusters, Coma being the most active, followed by Perseus, while Virgo is very relaxed. Typical turbulent velocities are a few $\approx100$\,km\,s$^{-1}$, very close to observed values. The resulting turbulent pressure support is $\approx1\%$ for Virgo, $\approx 6\%$ for Perseus, and $\approx 8\%$ for Coma within the central $1-2\%$ of $R_{200}$.
Compared to previous simulations and observations, measured velocities and turbulent pressure support are on average lower, in line with XRISM findings, thus indicating the importance of selection effects.

\end{abstract}

\keywords{Galaxy clusters --- Coma Cluster --- Virgo Cluster --- Perseus Cluster --- Intracluster medium --- Hydrodynamical simulations}

\section{Introduction}

Gas motions on various scales shape the structure of the Intra Cluster Medium (ICM), ranging from bulk motions and merger shocks on large scales (i.e. Mpc) to turbulence on small scales ($\mathcal{O}(10)\,$kpc), connected via a turbulent cascade \citep{Kravtsov&Borgani2012,Mohapatra+2021}.
They act as non-thermal pressure, leading to deviations from the assumption of hydrostatic equilibrium and to the hydrostatic mass bias \citep{Rasia+2006,Shi+2016,Biffi+2016,Vazza+2018a,Angelinelli+2020,Ettori&Eckert2022}.

Direct access not only to bulk-motions, but also to small-scale velocitiy dispersions by observations has been achieved by the \citet{HitomiCollaboration+2016,HitomiCollaboration+2018}, studying the broadening of X-ray spectral emission lines in the Perseus cluster. The velocity dispersion is mostly attributed to turbulence.
They find small-scale velocities of $100-200$\,km\,s$^{-1}$, corresponding to a turbulent pressure support of only $4\%$ in the central $60$\,kpc of the cluster.

Even deeper insight into the properties of gas dynamics and turbulence will be gained by the XRISM mission \citep{XRISMScienceTeam2022}, in particular by the Resolve X-ray micro-calorimeter \citep{Sato+2023}. One key scientific goal includes studying turbulence in the ICM.
Several observations have already been published, including the Centaurus cluster \citep{XRISMCollaboration+2025d}, Abell 2029 \citep{XRISMCollaboration+2025h,XRISMCollaboration+2025b}, Coma \citep{XRISMCollaboration+2025d}, Perseus \citep{Zhang+2025,XRISMCollaboration+2026a}, Virgo \citep{XRISMCollaboration+2026}, and Ophiuchus \citep{Fujita+2025a}. Even more observations have already been carried out and will be published in the near future.
As the XRISM mission progresses, it will provide a larger cluster sample, more pointings within individual clusters, and deeper observations for some of them.
All the clusters analyzed so far yield turbulent pressure fractions of only a few percent, and thus lie on the lower end compared to non-thermal pressure estimates from previous observations \citep[e.g.,][]{Eckert+2019,Gatuzz+2023a} which provide an upper bound for the turbulent pressure measured by XRISM/Hitomi. Only a few observations reach values similarly low as XRISM/Hitomi \citep[compare, e.g.,][]{Werner+2016,Gatuzz+2023a}. Differences among observations with different instruments can be explained by instrumental effects and analysis techniques. Also, the radius within which values are measured matters \citep{Lebeau+2025,Groth+2025a}, and comparing values measured in the same spatial region can solve part of the discrepancies \citep{Heinrich+2024}.

To disentangle the effect of all the processes contributing to the gas dynamics and the amount of turbulence, it remains crucial to obtain a theoretical foundation.
Most velocity measurements based on simulations, however, have focused on statistical comparisons to observations \cite[e.g.][]{Vazza+2012,Groth+2025a}. Besides differences due to the treatment of physics in the simulations, the specific choice of clusters can have major effects on measured velocities.
\citet{Lau+2017} have shown that simulations can indeed reproduce Hitomi observations for Perseus, given that the cluster has not experienced major mergers in the recent past.
First predictions aiming specifically for the XRISM observations have been made by \citet{Truong+2024} using the TNG-Cluster set \citep{Nelson+2024}. They inferred median non-thermal pressure fractions of $\approx 8\%$ from mock XRISM observations, higher than XRISM findings. Overall, XRISM observations lead to consistently smaller turbulent pressure fractions than simulation averages. Possible explanations for these deviations include selection effects, but also missing physics in the simulations, particularly a prescription for physical viscosity. The weighting employed for averaging as well as projection effects can further impact results and explain differences \citep{Vazza&Brunetti2026}.

In this work, we provide an even more direct comparison based on simulated local Universe galaxy cluster analogs from the SLOW simulations \citep{Dolag+2023,Seidel+2025}. These clusters have been shown to have excellent agreement regarding dynamical properties, formation histories, and thermodynamic profiles \citep{Hernandez-Martinez+2024}. In particular, we focus on three clusters -- Coma, Virgo, and Perseus -- that have strong constraints, thus are known to have good agreement with observations, and have already been observed by XRISM.

We use these clusters to demonstrate how the SLOW simulations provide exquisite simulated counterparts to XRISM observations, without the need to consider selection effects due to the currently small cluster set observed by XRISM. In particular, we derive turbulent velocity and pressure profiles, and compare our predictions to XRISM observations. We focus on the non-radiative simulations, including only gravity and hydrodynamics. This gives insight into the cosmological contributions in the form of mergers, accretion, and substructures on the amount of turbulence.
A more detailed analysis of the impact of astrophysical processes on turbulence will be performed with simulations including feedback and more individual clusters that will be published in follow-up studies.
Ultimately, the comparison between predictions from our simulations and XRISM observations will improve our understanding of the turbulent cascade in the ICM, give access to a better understanding of energy seeding by feedback events of a central AGN, and even constrain plasma properties such as viscosity.

This work is structured as follows. In Sec.~\ref{sec:simulations}, we describe the constrained initial conditions and the simulation code. We continue with the results in Sec.~\ref{sec:results}, and conclude and discuss our findings in Sec.~\ref{sec:conclusions}, which includes an outlook to possible future work.

\section{The Simulations} \label{sec:simulations}
\subsection{Local Universe Cluster Analogs} \label{sec:cluster_analogs}
We use the SLOW (Simulating the LOcal Web) simulations \citep{Dolag+2023}, which are based on a realization of the CLONES simulation set \citep{Sorce2018,Sorce+2021}.
Galaxy clusters have been cross-identified with observed clusters from different surveys based on positions, masses, and X-ray observables. General properties as well as radial profiles agree very well between observed clusters and simulated counterparts \citep{Hernandez-Martinez+2024}. For several of these clusters, zoom-in regions have been constructed \citep[B. A. Seidel et al. in prep, see also][]{Seidel+2025,Sorce+2021}.

In this study, we focus on three of these clusters -- Coma, Virgo, and Perseus -- which are among the first clusters observed by XRISM, while simultaneously being very well matched in the simulations. We choose these three clusters due to the good constraint quality in their vicinities, with several hundred independent constraints in each region.
Indeed, \citet{Hernandez-Martinez+2024} find that the proximity of these clusters to their observed positions and masses is very likely a result of the constraints. The probability of these clusters at given mass and distance to arise within a random simulation is only $\log_{10} P_{M_{500}}(r<\left|r_{\rm obs}-r_{\rm sim}\right|)=-3.5^{+0.1}_{-0.4}$ for Virgo, $-1.6^{+0.1}_{-0.8}$ for Coma, and $-1.6^{+0.6}_{-0.2}$ for Perseus.

Specific features such as substructures, bridges, and merger tracers found in X-ray observations have been shown to be very well reproduced within the simulated SLOW counterparts for different clusters \citep[see,e.g.,][]{Olchanski&Sorce2018,Sorce+2021,Dietl+2024,Reiprich+2025a}.

The observer position within the cosmological box has been chosen to reproduce the observed projected positions of the galaxy clusters on the sky.
Virgo, being very close to us, uses an alternative box center, which improves its individual position.
All simulations adopt a background cosmology according to \citet{PlanckCollaboration+2014} with $H_0=67.77$\,km\,s$^{-1}$\,Mpc$^{-1}$, $\Omega_{\rm m}=0.307115$, $\Omega_{\Lambda}=0.692885$, and $\Omega_{\rm b}=0.0480217$.

\subsection{\textsc{OpenGadget3} Simulation Code} \label{sec:OpenGadget3}
The galaxy cluster analogs have been simulated with the cosmological SPH/MFM TreePM code \textsc{OpenGadget3} \citep[OpenGadget3 collaboration in prep., refer to][and references therein]{Groth+2023b}. 
The code was originally based on \textsc{Gadget-2} \citep{Springel2005} with several updates, including an updated treewalk and domain decomposition \citep{Ragagnin+2016}.

Long-range forces are calculated using a PMGrid at resolution $512^3$. An additional, high-resolution PMGrid of $1024^3$ cells is inserted around the high-resolution region, enlarged by a factor $1.1$.
Hydrodynamical accelerations are calculated utilizing the SPH method with $295$ neighbors and a Wendland C6 kernel \citep{Wendland1995,Dehnen&Aly2012,Donnert+2013}. A non-local timestep criterion \citep[``wakeup'',][]{Pakmor+2012} ensures the stability around strong shocks.
Time-dependent artificial viscosity and conductivity are calculated based on second-order accurate gradient estimates \citep{Price2012b,Beck+2016}.
Physical conduction including a description for saturation \citep{Jubelgas+2004,Dolag+2004} is applied in addition to the aforementioned artificial conductivity. The physical conductivity implementation is based on the solver described by \citet{Petkova&Springel2009}.

We do not use any additional subgrid models for cooling and feedback in this study. This allows for a first, cleaner comparison, but also adds limitations to the interpretation of results. The effects of these models will be discussed in follow-up studies.

The main halo, its center, and size are identified using \textsc{Subfind} \citep{Springel+2001,Dolag+2009}.

\section{Results} \label{sec:results}
\subsection{Dynamical States}
\begin{figure}
    \centering
    \includegraphics[width=\linewidth]{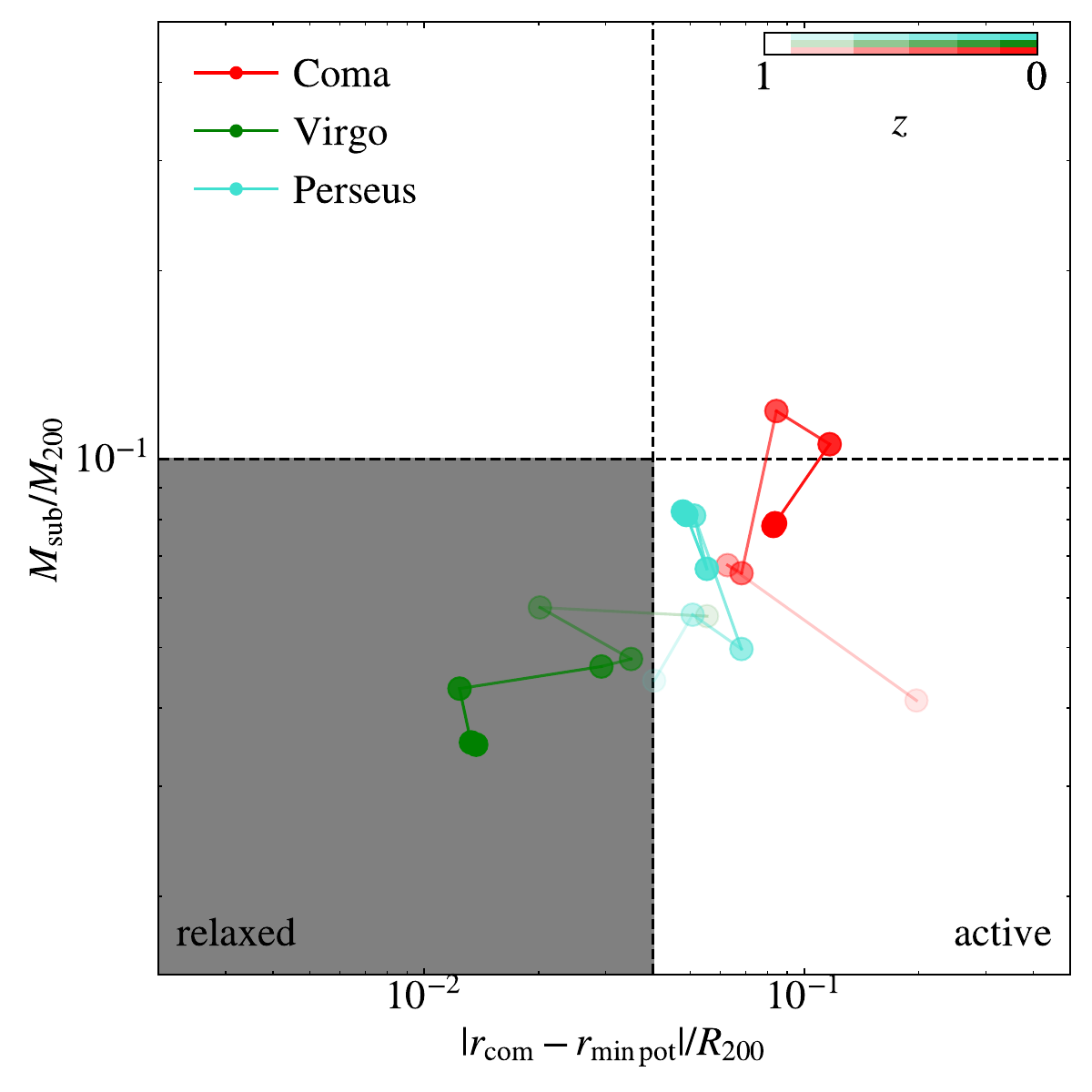}
    \includegraphics[width=\linewidth]{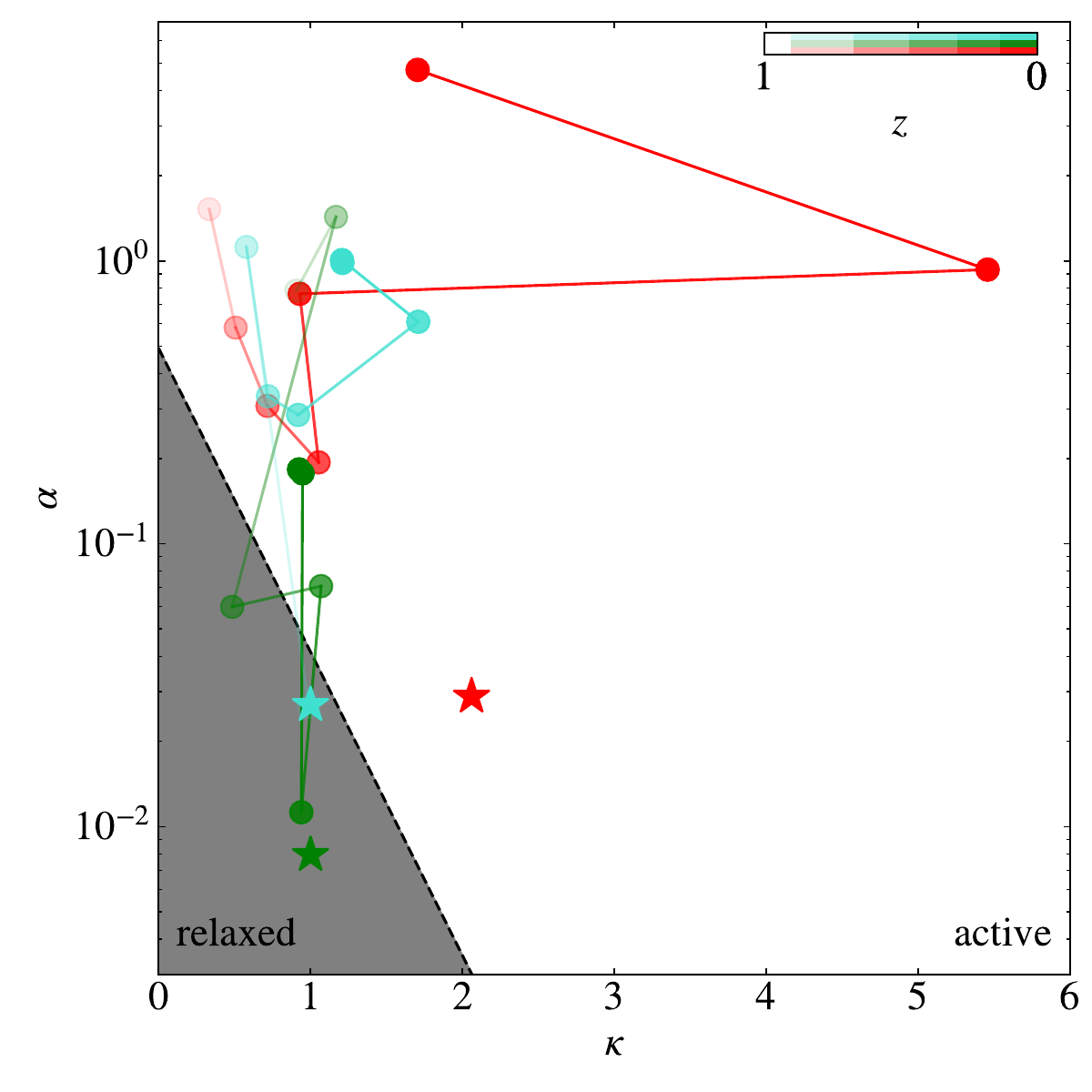}
    \caption{Relaxation criteria according to \citet{Cui+2017,Cui+2018} (top), and \citet{Yuan&Han2020} (bottom) for simulated Coma, Virgo, and Perseus analogs. The points indicate the redshifts of snapshots, for which data are derived. The connecting lines are meant to guide the eye, but do not contain physical information.
    The opacity of the lines/points indicates the evolution from $z=1.0$ (low opacity) to $z=0$ (high opacity).
    The dashed lines denote the thresholds for each criterion. Relaxed clusters lie within the gray area. Observed values by \citet{Yuan+2022} are shown as stars for comparison.}
    \label{fig:dynamical_state}
\end{figure}

We use two simulation-based criteria to classify the dynamical state and distinguish active from relaxed clusters, as described by \citet{Groth+2025a} and based on \citet{Cui+2017,Cui+2018}. Clusters are considered relaxed if the offset of the center of mass $r_{\rm com}$ compared to the position of the minimum potential $r_{\rm min\,pot}$ is less than $0.04~$R$_{200{\rm m}}$ and the mass enclosed in substructures $M_{\rm sub}$ does not exceed $0.1~$M$_{200{\rm m}}$. They are classified as active if one of the aforementioned criteria is not met. In Fig.~\ref{fig:dynamical_state}, we illustrate the evolution of both criteria with redshift for our simulated cluster analogs.

Observations do not have direct access to 3D mass distributions but determine the dynamical state based on the 2D-distribution of the X-ray surface brightness. 
To enable a closer comparison with observations, we also include two observation-based criteria, the asymmetry factor $\alpha$, describing the flux differences among symmetry pixels, and the profile parameter $\kappa=(1+\epsilon)/\beta$ combining the information content of the ellipticity $\epsilon$ and power-law index $\beta$, both evaluated within a $500$\,kpc aperture, as defined by \citet{Yuan&Han2020}, shown in the same figure. The combined morphology index $\delta=0.68\log_{10}\alpha+0.73\kappa+0.21$ provides a very good discrimination between relaxed ($\delta<0$) and active clusters ($\delta>0$).
The 2D mock X-ray maps that these measurements are based on were generated with SMAC \citep{Dolag+2005}, as described in App.~\ref{app:xray_images}.
Images are smoothed with a Gaussian filter as described by \citet{Yuan&Han2020}.

Overall, the level of cluster activity or relaxation matches well between the two criteria and also observations.
Simulation-based criteria show that Virgo is a relaxed cluster, as also predicted from observations based on the cool core even though mild sloshing motions are present \citep{Gatuzz+2022}. Perseus, which shows a cool core but sloshing motions in observations \citep{Fabian+2006}, is classified as active, but very close to being relaxed, while Coma, which shows traces of a recent merger with large substructures and two BCGs \citep{Neumann+2003}, is the most active cluster of the three.

The observation-based criteria, however, do not lead to any cluster classified as relaxed at $z=0$. This is a consequence of missing cooling and feedback in our simulation setups, as both of the processes strongly affect the central density profile. In particular, cooling can lead to colder and more compact cluster cores, reducing the profile parameter, bringing values closer to observations by \citet{Yuan+2022}. Feedback, in contrast, can lead to smoother large-scale gas distributions, reducing the asymmetry factor, overall pushing values to a lower morphology index.

\subsection{Turbulent Pressure Profiles}
The general analysis of the turbulent pressure profiles follows \citet{Groth+2025a}. The turbulent pressure is calculated based on the multi-scale filtered velocity \citep{Vazza+2012,Valles-Perez+2021a}, obtained using the \textsc{vortex-p} code \citep{Valles-Perez+2024}. The filtering length varies between $30$\,$h^{-1}$kpc and $1000$\,$h^{-1}$kpc depending on the region.
One main difference is the usage of volume-weighting instead of mass weighting, both within the multi-scale filtering and the non-thermal pressure averaging. This ensures more weight goes to the diffuse volume-filling gas, similar to observations. In addition, we exclude cold gas of temperatures $T < 10^5\,{\rm K}$ from the non-thermal pressure calculation which would not be visible in X-ray to bring the comparison even closer to observations. For non-radiative simulations, however, this exclusion of colder gas does not strongly affect results.
Finally, we use spherical annuli instead of elliptical shells.

The resulting, multi-scale filtered velocity generally corresponds to the turbulent velocity, filtering out bulk motions on larger scales. Deviations from this assumption are possible for strongly stratified clusters, where the strong gradients can impact the local filtering scale. In such cases, also larger-scale motions that are filtered out can be part of the turbulent cascade. As we did not find strong variations of the smoothing scale with radius, in our analysis the assumption of multi-scale filtered velocities corresponding to turbulence appears to be satisfied.

\begin{figure}
    \centering
    \includegraphics[width=\linewidth]{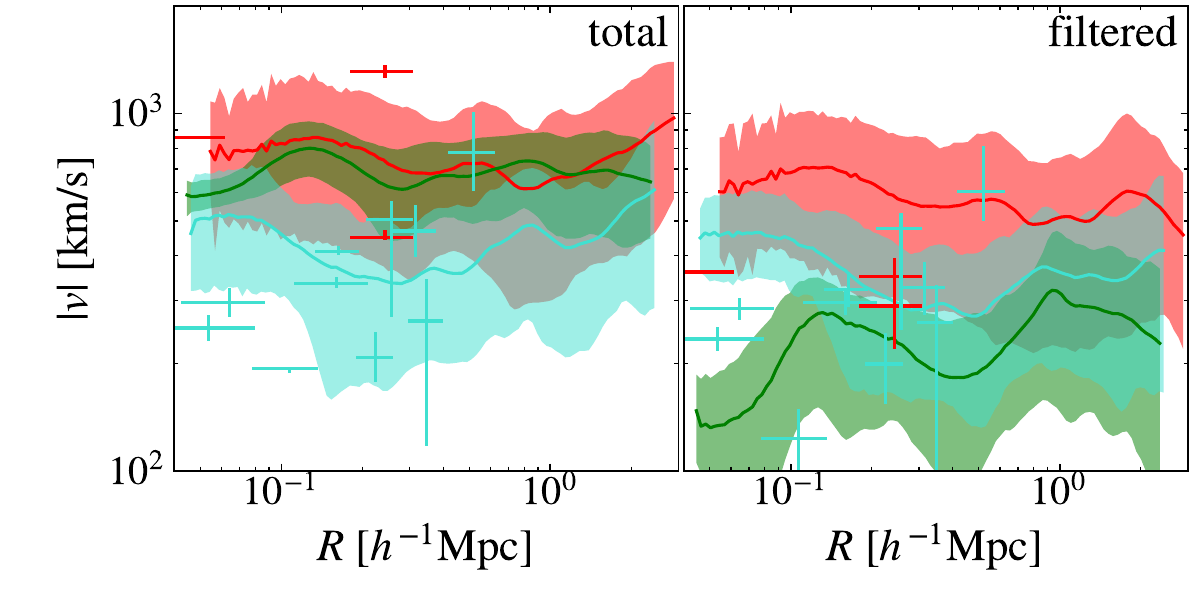}
    \caption{Radial total and multi-scale filtered, turbulent velocity profiles at redshift $z=0$ (solid lines). Same colors as in Fig.~\ref{fig:dynamical_state}. Observed velocity dispersions (right) and combined dispersions and bulk-motions (left) by XRISM/Hitomi (cyan) and XRISM (red) are overplotted as a comparison (straight lines with variance).}
    \label{fig:velocity_profiles}
\end{figure}

Total 3D velocities are measured relative to the BCG velocity, which we approximate by the mean DM velocity inside $50$\,$h^{-1}$kpc around the \textsc{Subfind} center.
The resulting velocity profiles are shown in Fig.~\ref{fig:velocity_profiles}.
Comparison values measured by Hitomi/XRISM shown in the total velocity panel are quadratically combined large-scale (bulk) motions and small-scale velocity dispersions $v_{\rm total}^2=v_{\rm bulk}^2+\sigma_v^2$.
This combination assumes isotropic turbulence and bulk motions, as also done within observation analysis.

We note that when ranking the clusters according to multi-scale filtered, turbulent velocities, the same ranking also applies to their dynamical state, showing a clear impact of the dynamical state on the turbulent velocity, and a slightly less pronounced impact on the bulk velocity.
The total velocity reaches up to $800$\,km\,s$^{-1}$ for Coma. Virgo has a total central velocity of $600$\,km\,s$^{-1}$, while Perseus only reaches $\approx 400-500$\,km\,s$^{-1}$.

Filtered velocities are significantly smaller, emphasizing the necessity to consider the filtering when comparing different studies. Coma yields the highest filtered velocities of $\approx 600$\,km\,s$^{-1}$ in the center, consistent with \citet{XRISMCollaboration+2025d}. Perseus reaches $\approx 500$\,km\,s$^{-1}$, and within the uncertainty range consistent with the \citet{HitomiCollaboration+2016,HitomiCollaboration+2018} results even though they tend to be higher.
Virgo yields the smallest turbulent velocities of $\approx 100-200$\,km\,s$^{-1}$ in the center.

\begin{figure}
    \centering
    \includegraphics[width=\linewidth]{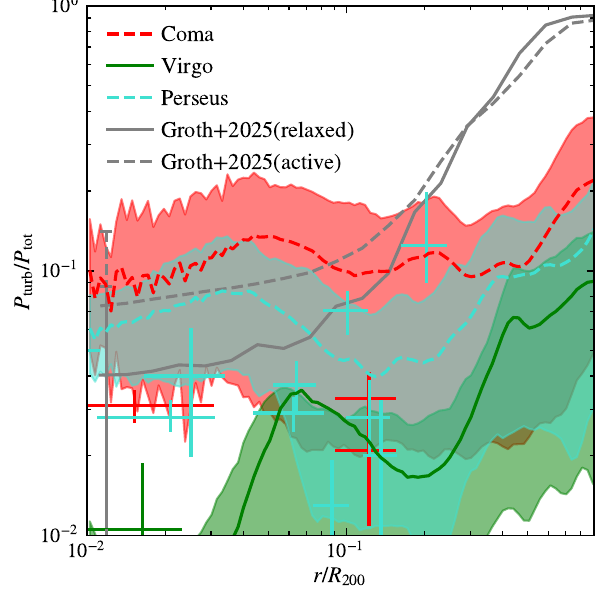}
    \caption{Radial turbulent pressure profile of all clusters at redshift $z=0$. Same colors as in Fig.~\ref{fig:dynamical_state}. Active clusters are shown with a dashed line, while relaxed clusters are indicated by a solid line.
    Observed turbulent pressure fractions by Hitomi (cyan) and XRISM (red) are overplotted as a comparison. In addition, mean values for relaxed (solid black) and active (daßshed black) clusters by \citet{Groth+2025a} obtained with the multi-scale filtering are shown, including the central $1\sigma$ scatter as an errorbar.}
    \label{fig:turbulent_pressure_profiles}
\end{figure}

The turbulent pressure profile calculated from the multi-scale filtered velocities is shown in Fig.~\ref{fig:turbulent_pressure_profiles}.
Central values for Virgo and Perseus are smaller than the average values found by \citet{Groth+2025a}, and even Coma, a very active cluster, is only marginally larger than statistical averages.
This gives a hint that selection bias can indeed be part of the explanation for the very low non-thermal pressure found by XRISM.

Overall, excellent agreement between Hitomi/XRISM observations and simulated counterparts is found for all clusters in the central region assuming the velocity dispersion accounts for all turbulent motions. Nevertheless, this comparison is also affected by methodological differences. As for Coma, values are higher further out in the simulation counterparts by a factor $\approx 2$, outer regions being also more prone to timing differences and to the presence of specific substructures, as visible in the increased uncertainty range both for simulations and observations.

\subsection{Mock X-Ray Spectra}
To bring the comparison even closer to observations, we can use the X-ray photon simulator \textsc{Phox} developed by \citet{Biffi+2012,Biffi+2018,Vladutescu-Zopp+2023}.
The emissivity of the gas is calculated based on the APEC\footnote{v3.0.9} model by \cite{Smith+2001}, including continuum Bremsstrahlung and line emission, leveraging the XSPEC implementation \citep{Arnaud1996}. 
Given the non-radiative nature of the simulations, a constant metallicity value for all gas particles is adopted equal to $Z=0.3Z_{\odot}$ with respect to the solar reference values by \citet{Anders&Grevesse1989}. 
We use \textsc{Phox} units 1 and 2, creating a photon list for an effective area of $210$\,cm$^2$, and clusters positioned at observed redshifts. Observation times of $622$\,ks for Coma, $230$\,ks for Perseus, and $33$\,ks for Virgo were used to obtain similar photon counts within the selected field-of-view for all three clusters.
The final spectrum is scaled according to the effective area at each energy, including the effect of the closed gate valve\footnote{The effective area was taken from the ARF file in SIXTE (\url{https://www.sternwarte.uni-erlangen.de/sixte/instruments/})}. 
In this work, we do not include other instrumental effects such as a PSF and the instrumental energy response.

\begin{figure}
    \centering
    \includegraphics[width=0.99\linewidth]{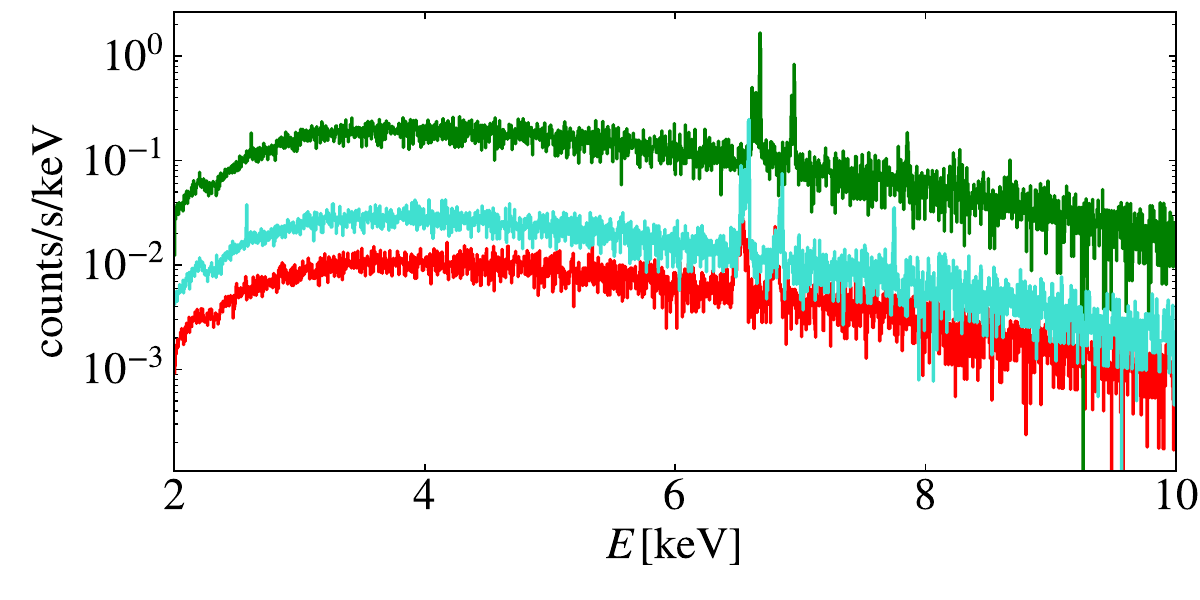}
    \includegraphics[width=0.99\linewidth]{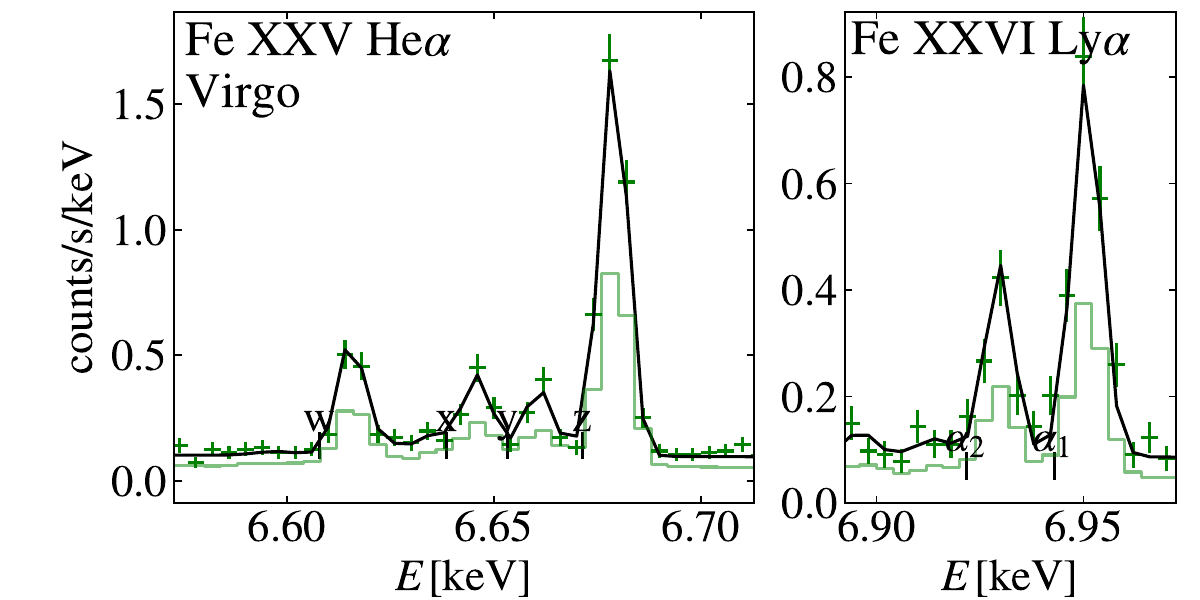}
    \includegraphics[width=0.99\linewidth]{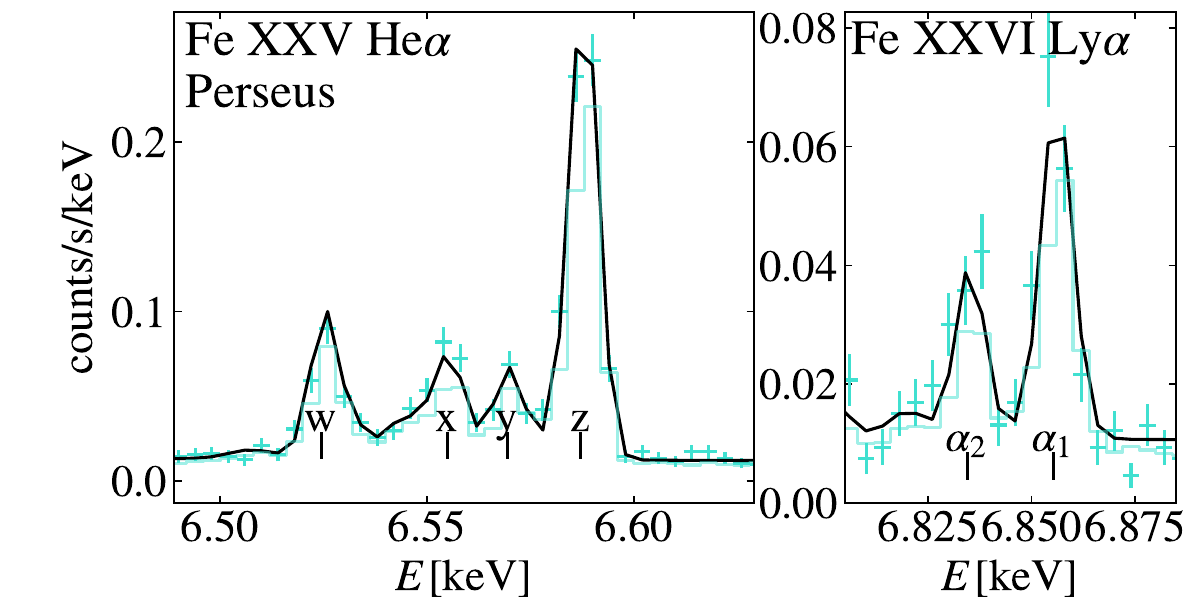}
    \includegraphics[width=0.99\linewidth]{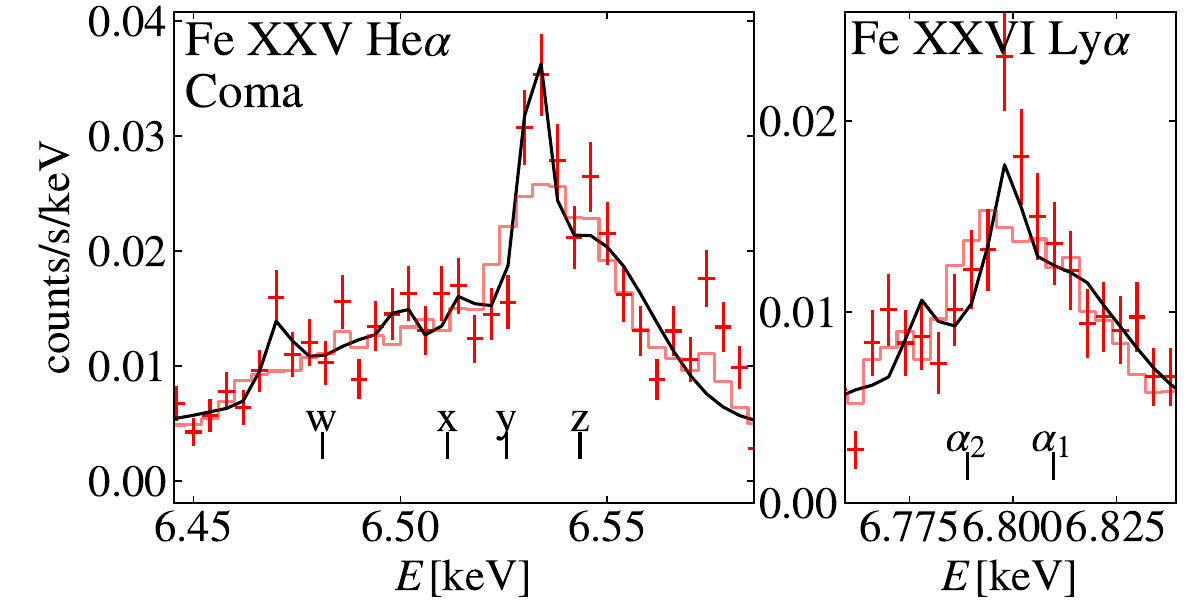}
    \caption{Broadband mock X-ray spectrum (top) and iron He$\alpha$ and Ly$\alpha$ line complexes (bottom) simulated with \textsc{Phox}. Same colors as in Fig.~\ref{fig:dynamical_state}.
    The position of the spectral lines shifted only due to the mean redshift of the cluster galaxies is shown as vertical lines. At low opacity as a solid line, we overplot the spectra at larger filtering length corresponding to $z=0.08$, scaled down to similar total emission. The black solid line shows the XSPEC fit of the spectrum.}
    \label{fig:line_profiles}
\end{figure}

Photons are shifted in energy due to the expansion of the Universe and gas line-of-sight velocity, leading to line shifts and broadening according to gas motions on different scales.
We collect all photons within an opening angle of $3\,$arcmin, as for XRISM.
Resulting mock X-ray spectra, including a broad-band spectrum as well as a zoom onto the Fe XXV He$\alpha$ and Fe XXVI Ly$\alpha$ line complexes for central pointings within all clusters are shown in Fig.~\ref{fig:line_profiles}.

The line profile of Coma is significantly broadened by several $10$\,eV. Virgo and Perseus, in contrast, show much narrower lines. Also, a shift of the line position can be observed, strongest for Coma, followed by Virgo, consistent with their larger bulk velocities in Fig.~\ref{fig:velocity_profiles}. Perseus, which has the lowest bulk velocity, shows hardly any line shift.

We perform a simple XSPEC\footnote{v12.12} fitting procedure on the retrieved spectra using a single temperature \texttt{bapec} model at XRISM-like energy resolution of 4 eV with a dummy response function. Overall, single temperature models for the Perseus and Virgo fields show great convergence, yielding small velocity dispersions of $\sigma_{z}=106\pm7\,\mathrm{km\,s}^{-1}$ and $\sigma_{z}=73\pm10\,\mathrm{km\,s}^{-1}$, respectively, in addition to thermal broadening, compared to the observational value of $\sigma_{z}=144\pm6$\,km\,s$^{-1}$ for Perseus \citep{HitomiCollaboration+2018} and $\sigma_{z}=262_{-46}^{+52}$ for Virgo \citep{XRISMCollaboration+2026}.
Our simulated values are consistently lower for these two clusters, solving the previously known tension for Virgo \citep{XRISMCollaboration+2025j}, and leaving room for effects from AGN feedback in the center of these clusters, which are not accounted for in our simulations.

For the Coma field, a two-temperature \texttt{bapec} ($\sigma_{z,1}=77\,\mathrm{km\,s}^{-1}$, $\sigma_{z,2}=668\,\mathrm{km\,s}^{-1}$) model was needed to account for a bimodal line-of-sight velocity structure in the chosen region caused by the interaction of the two BCG analogs.
This was not necessary for the Coma observations, for which \citet{XRISMCollaboration+2025d} could fit a single-component spectrum with $\sigma_{z}=208\pm12$\,km\,s$^{-1}$ and also resolve the individual lines more clearly, even though they overlap. Small timing and orientation differences, as well as differences in the precise choice of the region particularly relative to the two BCGs, can affect the measured velocity structure for this highly active system. A more detailed study of the variation of the spectrum depending on the aforementioned parameters would be required to fully understand the differences.

A key difference between the clusters is their velocity structure. Meanwhile, the choice of filtering lengths due to the different distances of the clusters also plays an important role. We study this effect by choosing a region size for all clusters as if they were located at an arbitrary redshift $z=0.08$ but scaling the emission back to the original area to have similar total emission. Virgo and Perseus, which are very relaxed, show no significant increase in turbulent line broadening.
Coma, in contrast, which is much more active, shows significantly broader lines by a factor of $\gtrsim 2$ due to the mixing of bulk and turbulent motions, which would thus lead to an overestimation of the amount of turbulent velocity and pressure.

This underlines the necessity to consider the physical filtering length when comparing clusters at different distances. The farther away a cluster is located, the larger the turbulent velocity estimate is.

\section{Discussion and Conclusions} \label{sec:conclusions}

In this paper, we used a constrained simulation as a comparison to XRISM and Hitomi observations. The properties of the simulated clusters show an excellent agreement with their observed counterparts. 
In particular, Coma is the most active cluster, followed by Perseus. Virgo is very relaxed, consistent with the very regular structure and cool core found in observations.

The difference in their dynamical state also manifests in the total and turbulent velocity profiles, and derived turbulent pressure fractions. Both for velocity and pressure profiles, Coma yields the highest values, followed by Perseus. Virgo, as the most relaxed cluster, shows the lowest amount of turbulence.
Notably, even for Coma -- the most active cluster -- the turbulent pressure support remains at only a few percent level. This result is in excellent agreement with XRISM observations assuming the velocity dispersion accounts for the turbulence. The most direct comparison can be done using mock spectra. Differences are visible in the mock-spectrum for Coma, in particular the required two-temperature model and the less clearly resolved lines. These are most likely related to timing differences and region matching.
Measured velocities for Virgo and Perseus from simulations are a few $100$\,km\,s$^{-1}$, slightly below  Hitomi/XRISM findings which can potentially be accounted for by currently ongoing kinetic feedback from the AGN \citep[compare, e.g.,][]{XRISMCollaboration+2025j}. This will require more detailed follow-up simulations to be tested.

Overall, the values are lower than predicted by other simulations that performed a more statistical comparison, independent whether they included feedback processes or not. This remains true even when trying to restrict the analysis to clusters of similar mass and inner properties as performed by \citet{Truong+2024}. 

The highly subsonic turbulence detected in all three clusters, which is yet consistent with the constrained simulation, highlights that selection effects may partially explain the low non-thermal pressure fractions observed by XRISM.
The evolutionary history is a key ingredient driving the amount of turbulence.
Our findings imply that constrained simulations offer a unique path when comparing to observations, independent of selection effects.

We provide mock spectra for all clusters, showing similar amounts of line broadening as found by XRISM. Consistent with derived velocities, Coma shows the broadest lines, while Perseus and Virgo show significantly more peaked line profiles. Besides the velocity structure, the physical filtering length becomes critical when interpreting XRISM measurements of clusters located at different distances. The effectively lower spatial resolution for clusters farther away leads to bulk motion being misinterpreted as turbulence.

In later, more detailed studies, cooling and feedback processes should be included in the simulations, which could impact the amount of turbulence, potentially increasing turbulent velocities. As \citet{Hernandez-Martinez+2024} found good agreement also in full-physics runs regarding thermodynamic properties, we expect these comparisons to be even more realistic. Further effects can arise due to the impact of plasma properties, in particular viscosity, which can potentially reduce the amount of turbulence. Further more detailed studies of a larger number of clusters will be necessary to find if there are statistical trends between the simulations, and whether modeling requires more physical processes to be consistent with XRISM.

Additional uncertainties in the amount of turbulence derived from our simulations can occur due to timing differences, which could be overcome by analyzing the simulation at the observational redshift of each
cluster, or even studying the temporal evolution close to the observed redshift, finding the best-fitting redshift.
In addition, more precise matching of the XRISM regions to observations would give insight into the possible amount of uncertainty within the outskirts.
Finally, line-of-sight and center differences can be explored, using alternative choices based on substructure or the nearby cosmic web.

Comparisons to observations could be improved even further by matching the filtering length instead of performing multi-scale filtering, and also by using instrumental effects.

Our work shows that constrained simulations provide a unique opportunity for one-to-one comparisons with observations due to the matching evolutionary history, dynamical and thermodynamic properties. Ultimately, this will allow for even tighter constraints regarding sub-resolution feedback models and plasma properties by comparing the velocity structure of a larger set of clusters between the SLOW simulation and XRISM observations.
In future studies, we plan to analyze more clusters in greater detail, exploiting the full predictive power of the SLOW simulation.

\section*{Data Availability}

We make all relevant derived data to reproduce the plots available online. They can be downloaded as an HDF5 file, which also contains relevant meta-information, at \url{https://doi.org/10.5281/zenodo.16778250}. Original simulation data and the simulation code \textsc{OpenGadget3} will be shared upon reasonable request.

\section*{Acknowledgments}
We thank the anonymous referee for the constructive feedback which helped to improve the quality of this paper.
We thank J. Stoiber for discussions on how to determine the north direction based on the large-scale structure environment.

FG, KD and BAS acknowledge support by the COMPLEX project from the European Research Council (ERC) under the European Union’s Horizon 2020 research and innovation program grant agreement ERC-2019-AdG 882679.
FG and KD acknowledges support by the Deutsche Forschungsgemeinschaft (DFG, German Research Foundation) under Germany’s Excellence Strategy - EXC-2094 - 390783311.

KD, JS and BAS acknowledge support by the grant agreements ANR-21-CE31-0019 / 490702358 from the French Agence Nationale de la Recherche / DFG for the LOCALIZATION project.

MV is supported by the Fondazione ICSC National Recovery and Resilience Plan (PNRR), Project ID CN-00000013 "Italian Research Center on High-Performance Computing, Big Data and Quantum Computing" funded by MUR - Next Generation EU. MV also acknowledges partial financial support from the INFN Indark Grant.

SVZ acknowledges support by the Deut\-sche For\-schungs\-ge\-mein\-schaft, DFG\/ project nr. 415510302.

JS acknowledges support from Univ. Lille, UNIVERSITWINS project (Initiative d'excellence).

The simulations were carried out at the Leibniz Supercomputer Center (LRZ) under the project pn68na (CLUES).

\section*{Author Contributions}
FG was responsible for performing the analysis, writing, and submitting the manuscript.
MV contributed to the interpretation of the results and edited the manuscript.
BAS created the zoom-in initial conditions for the simulated regions.
SVZ helped with the Phox analysis, fitted the spectra, and edited the manuscript.
VB provided access to the Phox code.
KD obtained the funding, ran the simulations, and contributed to the interpretation of the results.
JS created the constrained initial conditions of the original simulation box.

\paragraph{Facilities}Gauss:LRZ, SuperMUC-NG2 Friendly User Phase

\paragraph{Software}OpenGadget3 (OpenGadget3 collaboration in prep.), julia \citep{Bezanson+2014}, GadgetIO \citep{GadgetIO}, Phox  \citep{Biffi+2012,Biffi+2018,Vladutescu-Zopp+2023}, matplotlib \citep{Hunter2007}, SMAC \citep{Dolag+2005}, XSPEC \citep{Arnaud1996}

\begin{appendix}

\section{Mock X-Ray Images} \label{app:xray_images}

We create mock X-ray images in the energy band $0.5$ to $7$\,keV using emissivities based on tabulated cooling tables by \citet{Sutherland&Dopita1993} with SMAC \citep{Dolag+2005}.
The center is determined using \textsc{Subfind} and the observer position based on the optimum projected cluster positions in the simulation box.
The simulation box is in supergalactic coordinates with the $z$-direction pointing north. We use an alternative north direction than that of the simulation box.
Based on the three closest cross-identified clusters, we choose the direction that minimizes their angular deviation between observed and simulated positions relative to the cluster, while keeping the observer position fixed.
To allow for a more direct comparison to observations, we transform the coordinate system from supergalactic to J2000 coordinates.

\begin{figure*}
    \centering
    \includegraphics[width=0.99\textwidth]{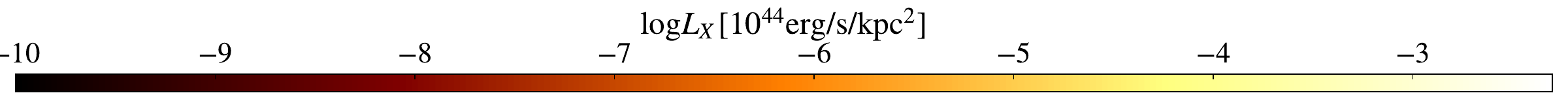}
    \includegraphics[width=0.32\linewidth]{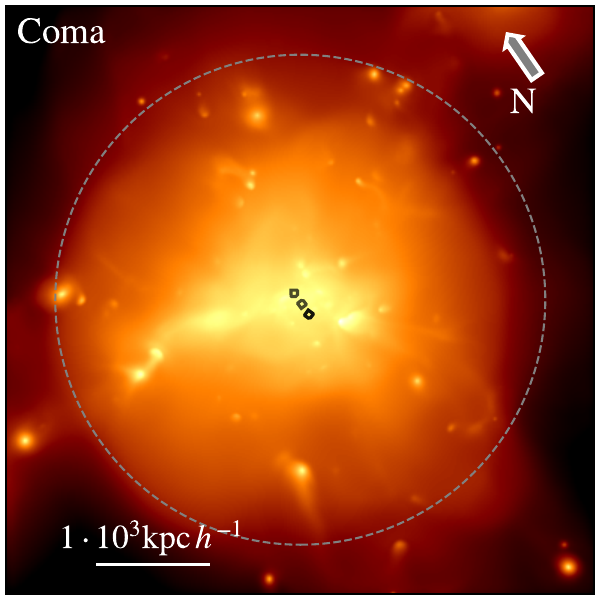}
    \includegraphics[width=0.32\linewidth]{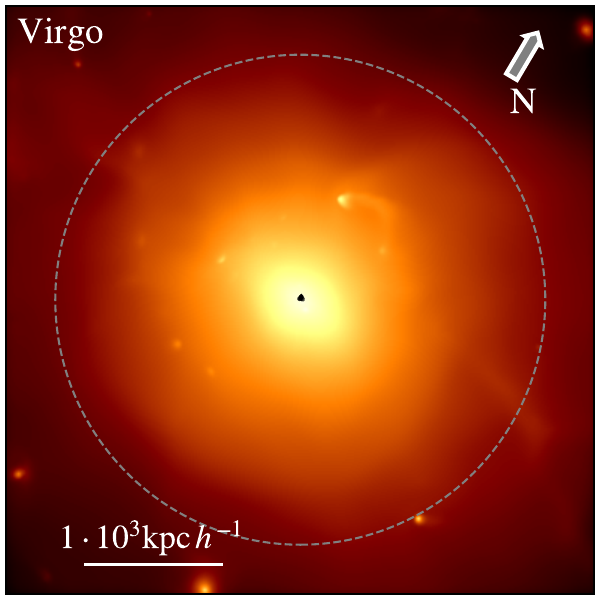}
    \includegraphics[width=0.32\linewidth]{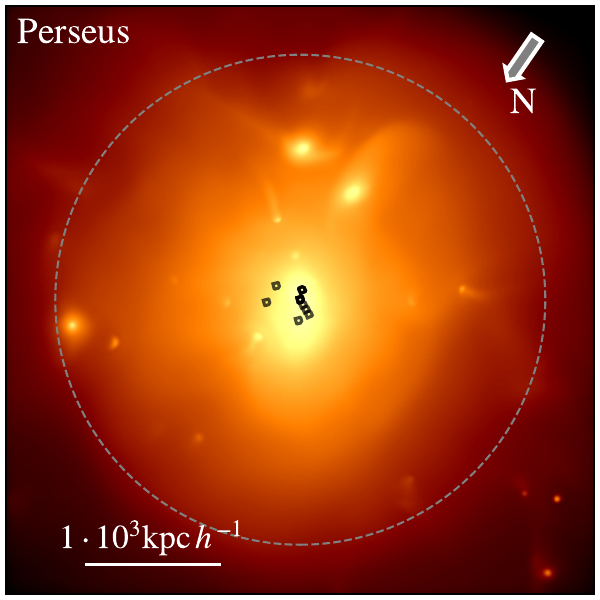}
    \caption{Mock X-ray images created with SMAC with tabulated cooling tables \citep{Sutherland&Dopita1993} in the energy band from $0.5$ to $7$\,keV. The dashed circle denotes $R_{\rm vir}$, and the arrow the north direction in J2000 coordinates. The black squares indicate the location of the XRISM pointings.}
    \label{fig:xray_images}
\end{figure*}

These X-ray mock images are shown in Fig.~\ref{fig:xray_images}.
Already from these images, and consistent with the findings discussed in the main text, Virgo is the most relaxed cluster with a very regular, symmetric density distribution. Perseus shows more substructures, but overall appears roughly spherical in the center. Coma is the most active cluster with many substructures and highly irregular central X-ray emission dominated by two dense substructures.
\citet{Hernandez-Martinez+2024} have already shown that the overall emissivity is close to observed values. For our non-radiative simulations, central values are typically smaller due to the missing cooling in simulations, which does not allow for the formation of a cool core.

For the calculation of dynamical state parameters, we apply an additional Gaussian smoothing at a scale of $10$\,kpc for Virgo and $30$\,kpc for Coma and Perseus, consistent with \citet{Yuan&Han2020,Yuan+2022}.

\end{appendix}

\bibliography{references}{}
\bibliographystyle{aasjournalv7}

\end{document}